\documentclass[twocolumn,secnumarabic,aps,pre,10pt]{revtex4}

\usepackage[english]{babel}
\usepackage{amssymb,color}
\usepackage{mathrsfs}
\usepackage{amsmath}
\usepackage[dvips]{graphicx}
\usepackage{epsfig,bm}

\newcommand{\bitem}{\begin{itemize}}
\newcommand{\fitem}{\end{itemize}}
\newcommand{\beq}{\begin{equation}}
\newcommand{\eeq}{\end{equation}}
\newcommand{\beqa}{\begin{equation} \begin{array}{rcl}}
\newcommand{\eeqa}{\end{array} \end{equation}}
\newcommand{\pt}{\partial}

\newcommand{\fb}{\bar{f}}
\newcommand{\dd}{\mbox{d}}
\newcommand{\fh}{\hat{f}}
\newcommand{\Vc}{\check{V}}

\begin{document}

\title{Vlasov equation for long-range interactions on a lattice}

\author{R.~Bachelard}
\email{bachelard.romain@gmail.com}
\affiliation{University of Nova Gorica, School of Applied Sciences, Vipavska 11c, SI-5270 Ajdovcina, Slovenia}

\author{T. Dauxois}
\email{Thierry.Dauxois@ens-lyon.fr}
\affiliation{Laboratoire de Physique de l'\'Ecole Normale Sup\'erieure de Lyon, Universit\'e de Lyon, CNRS, 46 All\'ee d'Italie, 69364 Lyon c\'edex 07, France}

\author{G. De Ninno}
\email{giovanni.deninno@elettra.trieste.it}
\affiliation{Sincrotrone Trieste, S.S. 14 km 163.5, Basovizza (Ts), Italy
\\ University of Nova Gorica, School of Applied Sciences, Vipavska 11c, SI-5270 Ajdovcina, Slovenia}

\author{S.~Ruffo}
\email{stefano.ruffo@gmail.com}
\affiliation{Dipartimento di Energetica ``Sergio Stecco'', Universit\`a di Firenze and INFN, 
via S. Marta 3, 50139 Firenze, Italy
\\ Laboratoire de Physique de l'\'Ecole Normale Sup\'erieure de Lyon, Universit\'e de Lyon, 
CNRS, 46 All\'ee d'Italie, 69364 Lyon c\'edex 07, France}

\author{F.~Staniscia}
\email{fabio.staniscia@elettra.trieste.it}
\affiliation{Sincrotrone Trieste, S.S. 14 km 163.5, Basovizza (Ts), Italy
\\ Dipartimento di Fisica, Universit\`a di Trieste, Italy}

\date{\today}

%\tableofcontents

\begin{abstract}
We show that, in the continuum limit, the dynamics of Hamiltonian systems defined on a lattice 
with long-range couplings is well described by the Vlasov equation. This equation can be
linearized around the homogeneous state and a dispersion relation, that depends explicitly 
on the Fourier modes of the lattice, can be derived. This allows one to compute the stability 
thresholds of the homogeneous state, which turn out to depend on the mode number.
When this state is unstable, the growth rates are also function of the mode number. Explicit
calculations are performed for the $\alpha$-HMF model with $0 \leq \alpha <1$, for which
the mean-field mode is always found to dominate the exponential growth. The theoretical
predictions are successfully compared with numerical simulations performed on a finite
lattice. 
\end{abstract}

\maketitle

\section{Introduction}

Long-range interactions are found in many different domains of physics: self-gravitating systems, 
unscreened Coulomb systems, plasmas, cold atoms, two-dimensional 
fluids~\cite{Campa,Leshouches1,Assisi,Leshouches2,Gupta}. 
Systems with long-range interactions have interesting properties at equilibrium, such as ensemble 
inequivalence \cite{Barre}. In this paper we will be concerned with their dynamical 
out--of--equilibrium behavior. 

In the continuum limit, where the number of particles, $N$, goes to infinity, the dynamics of these systems is well 
described by the {\it Vlasov equation}~\cite{Balescu,Nicholson}. It has been found that stable stationary states 
of this equation correspond to quasi-stationary states (QSS) of the finite-$N$
system~\cite{yamaguchi}. Long-range systems generically show a rapid 
evolution towards QSS, where the system remains trapped for a time that
increases algebraically with $N$. In the long time limit the system eventually relaxes to
Boltzmann-Gibbs equilibrium. This complex relaxation process was first described
in pioneering papers of Henon~\cite{Henon} and Lynden-Bell~\cite{Lyndenbell}. 

The Vlasov equation is usually derived for Hamiltonians of the form
\beq
\label{typical}
H=\sum_{j=1}^N \frac{\mathbf{p}_j^2}{2}+ \frac{1}{2} \sum_{j \neq k=1}^N V(|\mathbf{q}_j-\mathbf{q}_k|)~,
\eeq
where $\mathbf{q}_j$ denotes the position of the $j$-th particle in three-dimensional space,
$\mathbf{p}_j$ its conjugate momentum and $V(x)$ the potential function.

In this paper, we will show that systems defined on a one-dimensional lattice with $N$ sites 
can also be described by a Vlasov equation, if the coupling constant decays algebraically 
with the distance between two sites of the lattice. 

We consider Hamiltonians of the following form
\beq
H=\sum_{j=1}^N \frac{\mathbf{p}_j^2}{2} + \frac{1}{2\tilde{N}} \sum_{j,k=1}^N 
\frac{v(\mathbf{q}_j,\mathbf{q}_k)}{|x_j-x_k|^\alpha},
\label{eq:HNbody0}
\eeq
where $x_j=ja$ is the {\it fixed} coordinate of the $j$-th site on a one-dimensional lattice (the lattice
constant $a$ will be set to unity) and $\tilde{N}=\sum_{j=1}^N j^{-\alpha} \sim N^{1-\alpha}$, with
$0 \leq \alpha < 1$, is a normalization constant which makes the potential energy extensive 
in $N$~\cite{kac}. We choose periodic boundary conditions, $x_1=x_{N+1}$,$\mathbf{q}_1=\mathbf{q}_{N+1}$,
$\mathbf{p}_1=\mathbf{p}_{N+1}$.

The vector variable $\mathbf{q}_j$ has now a different physical interpretation:
it represents the internal degrees of freedom attached to site $j$. These degrees of freedom
are assumed to be still described by a Hamiltonian structure with a momentum 
vector $\mathbf{p}_j$ conjugate to $\mathbf{q}_j$. The dimension $D$ of these vectors is 
not the dimension of the physical space in which the motion of the particles takes place, 
but the dimension of the space of internal degrees freedom, so that it can take values different from $D=3$.

Models of this kind were introduced by Dyson~\cite{dyson1,dyson2}, Thouless~\cite{thouless} in the 60's with the aim of understanding phase transitions in one dimension, and were also connected to the physics of Kondo effect~\cite{anderson,hamann}. However, the internal degrees of freedom were discrete Ising variables while, following~\cite{Tsallis}, we here generalize the model to continuous variables representing, e.g., XY spins. Moreover, the addition of a kinetic term in the Hamiltonian allows us to treat the microcanonical dynamics.

To be concrete, let us consider the case in which we attach to site $j$ a two-dimensional 
vector pointing to the unit circle: $\mathbf{q}_j=(\cos q_j,\sin q_j)$, with $-\pi \leq q_j < \pi$ an
angle. This vector physically represents an XY rotor with conjugate angular momentum $p_j$ pointing 
towards the $Z$-direction. For example, let the function $v(x,y)$ be the scalar product of the two arguments 
$v(\mathbf{q}_j,\mathbf{q}_k)=-J \mathbf{q}_j \cdot \mathbf{q}_k=-J \cos (q_j-q_k)$, where the coupling 
constant $J$ can be positive (ferromagnetic coupling) or negative (antiferromagnetic
coupling). This coupling has been already considered in the literature~\cite{Tsallis,tamarit,campa00,campa03} 
and the corresponding model goes under the name of $\alpha$-HMF, where the acronym stands for 
Hamiltonian Mean Field. Indeed, when $\alpha=0$ the model reduces to the HMF 
model~\cite{inagaki,pichon,hmf95}: the lattice structure is removed in this limit and the 
rotors are coupled with equal strength.

Two kinds of results were obtained for the ferromagnetic $\alpha$-HMF model. On the one hand, it was shown analytically that 
the statistical equilibrium is the same as the one of the all-equally-coupled $\alpha=0$ case, both in the canonical 
and in the microcanonical ensemble~\cite{campa00,campa03}.
These results were anticipated by numerical simulations performed in the microcanical ensemble~\cite{Tsallis,tamarit}.
On the other hand, simulations performed in out-of-equilibrium conditions have shown that 
no significant difference is observed in the features of the QSS with respect to the $\alpha=0$ 
case~\cite{tamarit,vandenberg,turchi}. 
This raises the question of the full equivalence between the model with all-to-all coupling and the lattice 
model (see also \cite{Mori}).

We will here show that the Vlasov formalism can be actually generalized to long-range interactions on 
a lattice. In the $N \to \infty$ limit, each infinitesimal element of the lattice contains an infinity 
of sites, to which a {\it local distribution function} in phase-space $f(\mathbf{q},\mathbf{p};x,t)$, 
depending also on the spatial coordinate $x$ along the lattice, can be associated.
A Vlasov equation for this local distribution function can be then derived, and the long-range interaction 
determines a global coupling between these local distributions at different spatial locations. A linear stability 
analysis of a state which is both uniform in the internal degrees of freedom and along the lattice will be performed and the
stability thresholds will be determined analytically. When this state is unstable, different Fourier 
modes grow with different rates and the zero mode, i.e. the mean-field mode of the $\alpha=0$ model, 
is found to dominate the instability process, indicating that, indeed, the dynamics shows similar features
with the HMF model.  

This paper is organized as follows: Section~\ref{sec:eqVl} is devoted to the derivation of the Vlasov equation for long-range interacting systems 
on a lattice. In Section~\ref{sec:dispersion} we discuss the linearization around the uniform state and
we find a dispersion relation expressed in terms of lattice Fourier modes. In Section~\ref{sec:HMFalpha} 
we derive the stability thresholds and the growth rates of the Fourier modes for the $\alpha$-HMF model, which
are tested against numerical simulations performed with finite $N$ in Section~\ref{simulations}. Finally, 
in Section~\ref{conclusion} we draw some conclusions.

\section{Vlasov equation on a lattice\label{sec:eqVl}}

In order to simplify the derivation, we restrict to one-dimensional internal degrees of 
freedom attached to the lattice site $j$. These degrees of freedom are represented by the 
canonically conjugate pair $(q_j,p_j)$. The Hamiltonian is the following
\beq
H=\sum_j \frac{p_j^2}{2} +\frac{1}{2\tilde{N}}\sum_{j,k=1}^N \frac{v(q_j,q_k)}{|x_j-x_k|^\alpha},
\label{eq:HNbody}
\eeq
where $v(x,y)$ is symmetric in the two arguments. The equations of motion are then
\begin{eqnarray}
\dot{q}_j &=& p_j,\label{eq:qj}
\\ \dot{p}_j &=& -\frac{1}{\tilde{N}}\sum_{k} \frac{v'(q_j,q_k)}{|x_j-x_k|^\alpha}~,\label{eq:pj}
\end{eqnarray}
where the prime denotes the derivative with respect to one of the two arguments. 
Since we have chosen periodic boundary conditions, we use the closest distance convention, i.e. the 
distance $|x_j-x_k|$ actually corresponds to $\mbox{min}(|x_j-x_k|,1-(|x_j-x_k|))$. 

In the continuum limit, the lattice (which for periodic boundary conditions can be thought to be a circle) 
is densely filled with sites: each infinitesimal lattice element $dx$ contains a diverging number of 
sites as $N$ goes to infinity. Therefore, one can define in this limit a distribution function $f(q,p;x,t)$,
which depends both on the internal variables $(q,p)$ and on the coordinate $x$ along the lattice, which is
confined to the interval $[0,1]$.
In the continuum limit, the equations of motion become
\begin{eqnarray}
\dot{q} &=& p,\label{eq:q}\\ 
\dot{p} &=& -\kappa_\alpha\iiint f(q',p';x',t) \frac{v'(q,q')}{|x-x'|^\alpha}\dd q' \dd p' \dd x',
\label{eq:pjconv}
\end{eqnarray}
where $\kappa_\alpha$ is such that $1/\tilde{N}\xrightarrow[N \to \infty]{}\kappa_\alpha/N$. The latter 
limit can be shown to yield $\kappa_\alpha^{-1}=\int_{-1/2}^{+1/2} dx/|x|^\alpha$. Equation~(\ref{eq:pjconv}) 
naturally leads to the definition of the long-range potential 
\beq
V_x[f](q,t)=\kappa_\alpha\iiint \dd q' \dd p' \dd x' f(q',p';x',t) \frac{v(q,q')}{|x-x'|^\alpha}~,
\label{eq:Vxf}
\eeq
which determines the motion of the conjugate variables $(q,p)$ at point $x$. This definition allows 
us to write the following Vlasov equation
\beq
\frac{\partial f}{\partial t}+p \frac{\partial f}{\partial q}-V'_x[f](q,t) 
\frac{\partial f}{\partial p}=0, \label{eq:Vl}
\eeq
where the prime refers to $\partial /\partial q$. It is interesting to note that this equation also derives 
from the following Hamiltonian
\beq
H[f]=\iiint \dd q' \dd p' \dd x' f(q',p';x',t) \left(\frac{p'^2}{2} + \frac{1}{2} V_x[f](q') \right).
\label{eq:HVlasov}
\eeq
Thus, Eq.~(\ref{eq:Vl}) actually represents a continuum of Vlasov equations along the lattice, 
globally coupled through the potential~(\ref{eq:Vxf}). The absence of a spatial derivative 
$\pt_x$ in Eq.~(\ref{eq:Vl}) reminds us that the lattice sites are fixed.

More rigorously, one should follow a procedure that begins with the definition 
of the so-called empirical measure
\beq
f_{e}(q,p;x,t)=\frac{1}{N}\sum_j \delta(q-q_j(t)) \delta(p-p_j(t)) \delta(x-x_j)~,
\eeq
and then derive the Klimontovich equation as done in \cite{braun,neunzert}. The Vlasov
equation could be then derived using a convenient averaging method.

It is also important to stress that definition~(\ref{eq:Vxf}) is consistent only 
for $\alpha<1$, since the integral $\int dx'/|x-x'|^\alpha$ is convergent around $x$ only 
for such values of $\alpha$. For $\alpha=1$ the integral diverges logarithmically,
we will not consider this case.

Let us comment about the normalization of $f$, an issue linked to the density of sites 
along the lattice. Since we are here considering a homogeneous lattice, each infinitesimal element 
$dx$ contains the same number of particles. Hence the norm of $f(q,p;x,t)$ has 
to be the same at any position $x$
\beq
\iint f(q,p;x,t)\dd q\dd p=1.
\eeq
Yet, our formalism is also suitable for heterogeneous lattices with non-uniform density
of sites $\rho(x)$, for which the normalization is given by  
\beq
\iint f(q,p;x,t)\dd q\dd p=\rho(x).
\eeq
The density $\rho(x)$  enters as a weight when deriving the local potential $V_x[f]$.

Finally, we note that the generalization to higher-dimensional lattices and to higher dimensional
spaces of internal degrees of freedom is straightforward. 
For $D$-dimensional conjugate variables $(\bm{q},\bm{p})$ defined at position $\bm{r}$ on a $d$-dimensional 
lattice, the $1/|x_j-x_k|^\alpha$ coupling turns into  $1/|\bm{r}_j-\bm{r}_k|^\alpha$, and the 
distribution function $f(\bm{q},\bm{p};\bm{r},t)$ obeys a Vlasov equation at each site $\bm{r}$.

\section{The dispersion relation\label{sec:dispersion}}

As anticipated in the introduction, the emergence of a Vlasov equation in long-range systems 
is of particular importance for their dynamics, since it is responsible for the long-lasting 
out-of-equilibrium regimes for finite $N$, the so-called QSS. This phenomenon is however 
bounded to the existence of Vlasov {\it stationary stable} states. We shall now 
turn to determine the conditions for such states to be present for the Vlasov equation on
a lattice (\ref{eq:Vl}). Let us consider a stationary solution $f_0$ of Eq.~(\ref{eq:Vl})
\beq
p \pt_q f_0 -V'_x[f_0](q) \pt_p f_0 = 0.
\label{eq:stat}
\eeq
From the above relation, it appears that the motion of a test-particle at position $x$ can 
be derived from the following Hamiltonian
\beq
h_x(q,p)=\frac{p^2}{2}+V_x[f_0](q)~.
\label{eq:h}
\eeq
Following the observation that $h_x(q,p)$ is a conserved quantity for stationary solutions $f_0$, 
several authors~\cite{BinneyTremaine,bertin,kaljnas,camporeale,lin2001,chavanis07,barre10,us10} studied inhomogeneous Vlasov equilibria expressing 
the stability problem in action-angle variables. However, for the sake 
of simplicity, we will here focus on homogeneous stationary states, both in position $x$ along the
lattice and in the internal variable $q$. This implies that $f_0=f_0(p)$ and $V'_x[f_0](q)=0$ for 
all $x$. The stability of $f_0$ can be studied by considering a small perturbation of the 
stationary distribution
\begin{equation}
f(q,p;x,t)=f_0(p) +\delta f(q,p;x,t).
\end{equation}
Inserting the above expression into the Vlasov equation (\ref{eq:Vl}) and dropping 
the quadratic term in $\delta f$ leads to the linearized Vlasov equation
\begin{eqnarray}
\pt_t (\delta f) + p \pt_q (\delta f) -f'_0(p) V'_x[\delta f](q,t)=0.
\end{eqnarray}
If we now focus on an eigenmode $\delta f_t(q,p;x)=e^{\lambda t}\bar{f}(q,p;x)$ of the linearized 
dynamics, we get
\begin{equation}
(\lambda +p\partial_q) \bar{f}(q,p;x)- f'_0(p)V'_x[\bar{f}](q)=0,
\end{equation}
which can be rewritten as
\begin{equation}
\partial_q \left(e^{\lambda \frac{q}{p}}\bar{f}\right)-\frac{\ e^{\lambda \frac{q}{p}}}{p}f_0'(p) 
V'_x[\bar{f}](q)=0.
\end{equation}
The integration over $q$ yields the dispersion relation
\begin{equation}
\bar{f}(q,p;x)-f'_0(p)\frac{e^{-\lambda \frac{q}{p}}}{p}\int_{q_0}^{q}
e^{\lambda \frac{q'}{p}} V'_x[\bar{f}](q')\dd q'=0,\label{eq:disp3}
\end{equation}
where we have assumed the integration constant to be zero. The solution of this dispersion 
relation depends on the form of the local potential $V_x[f]$. If we restrict to the derivation of
the stability threshold, we can set $\lambda=0$ in the previous expression
and perform the integral over $q'$, which gives
\begin{equation}
\bar{f}(q,p;x)-\frac{f'_0(p)}{p} V_x[\bar{f}](q,p;x)=0.
\label{eq:stabthreshold}
\end{equation}
This is indeed a set of equations for each position $x$, all coupled together 
by the $V_x[\bar{f}]$ term.

Now, since we have considered a lattice with periodic boundary conditions, it is natural 
to introduce the Fourier operator
\beq
\mathscr{F}_k[\bar{f}](q,p)=\frac{1}{2\pi}\int \dd x\ e^{-2i\pi k x}\bar{f}(q,p;x)=
\fh_k(q,p),
\label{eq:Fourierop}
\eeq
and rewrite the distribution $\bar{f}$ as a sum of Fourier modes
\beq
\fb(q,p;x)=\sum_k \fh_k(q,p)e^{2i\pi k x}.
\label{eq:decFourier}
\eeq
The potential $V_x[\fb]$ exhibits the interesting property of being diagonal in Fourier 
space~\footnote{Remark that, regarding numerical simulations of such systems, this property 
also allows one to compute the potential in Fourier space, thus reducing the {\it a priori} 
$N^2$ complexity of the double sum to the $N \ln N$ one of the Fast Fourier Transform.}. 
Indeed, inserting the sum (\ref{eq:decFourier}) into the definition~(\ref{eq:Vxf}) results into
\begin{eqnarray}
V_x[\fb](q)=\kappa_\alpha\sum_k  & & \left( \int \dd x' \frac{e^{2i\pi k x}}{|x-x'|^\alpha}\right)\label{eq:Vxfbar}
\\ && \times \iint \dd q' \dd p' \fh_k(q',p') v(q,q').\nonumber
\end{eqnarray}
Recalling the definition of the nearest distance on the periodic lattice, the above integral 
over $x'$ is actually performed from $x-1/2$ to $x+1/2$, so that it is appropriate to introduce 
the change of variables $x'\rightarrow y=x'-x$. One gets
\begin{eqnarray}
\int_{x-1/2}^{x+1/2} \dd x' \frac{e^{2i\pi k x'}}{|x'-x|^\alpha}&=&e^{2i\pi k x}\int_{-1/2}^{+1/2}  
\dd y\frac{e^{2i\pi k y}}{|y|^\alpha} \nonumber\\
&=&e^{2i\pi k x}c_k(\alpha)\int_{-1/2}^{+1/2} \frac{\dd y}{|y|^\alpha},
\end{eqnarray}
where $c_k(\alpha)$ is defined as
\beq 
c_k(\alpha)=\kappa_\alpha\int_{-1/2}^{+1/2} \frac{e^{2i\pi k y}}{|y|^\alpha}\dd y=
\frac{\displaystyle \int_{-1/2}^{+1/2} \frac{e^{2i\pi k y}}{|y|^\alpha}\dd y}
{\displaystyle\int_{-1/2}^{+1/2} \frac{1}{|y|^\alpha}\dd y}.
\label{cik}
\eeq
Equation~(\ref{eq:Vxfbar}) leads us to define the {\it mean-field potential}~$\Vc(q)$ for a distribution $g(q,p)$, which does 
not depend neither on~$\alpha$ nor on the position on the lattice
\beq
\Vc[g](q)=\iint \dd q' \dd p' g(q',p') v(q,q')~.
\label{eq:Vnonlocal}
\eeq
Remark that in the limit $\alpha=0$, when the long-range potential $V_x$ in formula (\ref{eq:Vxf})
does not depend any more on the position $x$ along the lattice, one has 
$\Vc[\fh]=\iint \dd q' \dd p' f(q',p') v(q,q')$, exactly as in mean-field models. 
Equation~(\ref{eq:Vnonlocal}) allows us to rewrite the dispersion relation (\ref{eq:disp3}) as
\begin{eqnarray}
0&=&\sum_k \fh_k(q,p)e^{2i\pi k x}-\sum_{k'} c_{k'}(\alpha)e^{2i\pi k' x} \nonumber
\\ &&\hskip 1.3truecm \times f'_0(p)\frac{e^{-\lambda \frac{q}{p}}}{p}\int_{q_0}^{q}e^{\lambda \frac{q'}{p}} \Vc'[\fh_k](q')\dd q'.\label{eq:disp4}
\end{eqnarray}
Thus, applying the inverse Fourier transform leads to the following set of decoupled dispersion relations 
for each Fourier mode $\fh_k$
\beq
\fh_k(q,p)-c_k(\alpha) f'_0(p)\frac{e^{-\lambda_k \frac{q}{p}}}{p}\int_{q_0}^{q}
e^{\lambda_k \frac{q'}{p}} \Vc'[\fh_k](q')|\dd q'|=0.\label{eq:disp6}
\eeq
The quantity~$\lambda_k$ represents the eigenvalue associated to mode $k$. It is interesting to 
note that, although the solution of these equations will generically depend on the specific form 
of the potential $\Vc[f]$, the dependence on the specific Fourier mode only appears through the 
coefficient $c_k(\alpha)$. We shall see in the next Section that this allows us to derive the 
growth rate for all Fourier modes in the case of the $\alpha$-HMF model.

\section{Stability thresholds and growth rates for the $\alpha$-HMF model \label{sec:HMFalpha}}

As a test-bed for systems of rotators on a ring-lattice with long-range couplings, we now turn 
to analyzing the dispersion relations for each Fourier mode of the $\alpha$-HMF model \cite{Tsallis,tamarit,campa00,campa03}. 
The rotators are thus coupled through a two-body potential $v(q,q')=-\cos (q-q')$, which is tuned down 
by a decay over the distance along the lattice $1/|x-x'|^\alpha$. 
The potential $\Vc$, defined in Eq.~(\ref{eq:Vnonlocal}) reads
\beq
\Vc[\fh](q)=-\kappa_\alpha\left(M_x[\fh]\cos q+M_y[\fh]\sin q\right),
\eeq
where the magnetization $M[\fh]$ is defined as
\begin{eqnarray}
{\bf M}[\fh]&=& M_x[\fh]+i M_y[\fh]
\\ &=& \iint \dd q' \dd p'\ \fh(q',p') \cos q' +i\iint \dd q' \dd p'\ \fh(q',p') \sin q'. \nonumber
\end{eqnarray}
Equation~(\ref{eq:disp6}) turns into
\begin{eqnarray}
0&=&\fh_k(q,p)-c_k(\alpha) f'_0(p)\frac{e^{-\lambda_k \frac{q}{p}}}{p}\label{eq:disphmf1}
\\ && \times \int_{q_0}^{q}e^{\lambda_k \frac{q'}{p}} \left(M_x[\fh_k]\sin q'-M_y[\fh_k]\cos q'\right)|\dd q'|.\nonumber
\end{eqnarray}
This equation can be solved by multiplying both sides by either $\cos q$ or $\sin q$ and integrating 
over the phase-space $(q,p)$, which leads to a linear system of equations in $M_x[\fh_k]$ and $M_y[\fh_k]$
\begin{eqnarray}
M_x[\fh_k]\left(1-c_k(\alpha) I_{X,Y}^{\lambda_k}[f_0]\right)+M_y[\bar{f}]c_k(\alpha)I_{X,X}^{\lambda_k} [f_0]=0,\nonumber
\\ M_x[\fh_k] c_k(\alpha) I_{Y,Y}^{\lambda_k}[f_0]-M_y[\bar{f}]\left(1+c_k(\alpha) I_{Y,X}^{\lambda_k}[f_0]\right)=0,\label{eq:linsyst}
\end{eqnarray}
where $I_{X,Y}^{\lambda}[f_0]$ is defined, for a scalar $\lambda$ and two functions $X(q),Y(q)$ (here 
$X(q)=\cos q$ and $Y(q)=\sin q$), as
\begin{equation}
I_{X,Y}^\lambda[f_0]=\int \dd p \frac{f'_0(p)}{p}\oint \dd q\, e^{-\lambda \frac{q}{p}}X(q)\int_{q_0}^q \dd q'e^{\lambda \frac{q'}{p}}Y(q').\label{eq:IXY}
\end{equation}
Assuming that eigenmodes $\bar{f}$ have a non-zero magnetization, the system of equations~(\ref{eq:linsyst}) 
implies
\begin{eqnarray}
0&=&\left(1-c_k(\alpha)I_{X,Y}^{\lambda_k}[f_0]\right)\left(1+c_k(\alpha)I_{Y,X}^{\lambda_k}[f_0]\right)\nonumber
\\ && +c_k(\alpha)^2 I_{Y,Y}^{\lambda_k}[f_0]I_{X,X}^{\lambda_k} [f_0].
\label{eq:disprel1}
\end{eqnarray}
Using the following formulae
\begin{eqnarray}
\int^q \dd q'\  e^{{\lambda_k} \frac{q'}{p}}\sin q'&=& \frac{e^{{\lambda_k} \frac{q}{p}}}{1+\frac{{\lambda_k}^2}{p^2}}\left(\frac{{\lambda_k}}{p}\sin q-\cos q\right),
\\ \int^q \dd q'\ e^{{\lambda_k} \frac{q'}{p}}\cos q'&=& \frac{e^{{\lambda_k} \frac{q}{p}}}{1+\frac{{\lambda_k}^2}{p^2}}\left(\sin q +\frac{{\lambda_k}}{p}\cos q\right),
\end{eqnarray}
one can calculate the integrals (\ref{eq:IXY}), so that dispersion relation~(\ref{eq:disprel1}) 
can be finally rewritten as
\begin{eqnarray}
0&=&\left(1+\pi c_k(\alpha)\int \dd p \frac{f'_0(p)}{p\left(1+\frac{{\lambda_k}^2}{p^2}\right)} \right)^2 \nonumber
\\ &&+\left(\pi c_k(\alpha)\int \dd p \frac{f'_0(p)}{p^2\left(1+\frac{{\lambda_k}^2}{p^2}\right)}\right)^2.\label{eq:disphomo2}
\end{eqnarray}
Note that in this relation only the coefficient $c_k(\alpha)$ changes from one Fourier 
mode to another.

Let us now consider a {\it waterbag} distribution \cite{yamaguchi}
\begin{equation}
f_{0}(p)=\frac{1}{2\pi}\frac{1}{2\Delta p}\left(\Theta(p+\Delta p)-\Theta(p-\Delta p)\right),
\end{equation}
where $\Theta$ is the Heavyside function. It's first derivative is
\begin{equation}
f'_{0}(p)=\frac{1}{2\pi}\frac{1}{2\Delta p}\left(\delta(p+\Delta p)-\delta(p-\Delta p)\right)~,
\end{equation}
where $\delta$ is the Dirac delta function.
Since this distribution is symmetric in $p$, the last term in Eq.~(\ref{eq:disphomo2}) vanishes, 
and the dispersion relation simplifies into
\begin{eqnarray}
0&=&1+\pi c_k(\alpha)\int \dd p \frac{f'_0(p)}{p\left(1+\frac{{\lambda_k}^2}{p^2}\right)}\nonumber
\\ &=&1-\frac{c_k(\alpha)}{2\Delta p^2\left(1+\frac{{\lambda_k}^2}{\Delta p^2}\right)}.
\end{eqnarray}
Thus, the eigenvalue of the $k$-th Fourier mode is given by
\begin{equation}
\lambda_k=\sqrt{\frac{c_k(\alpha)}{2}-\Delta p^2}.\label{eq:disphomo3}
\end{equation}
Hence, mode $k$ is stable provided $\Delta p^2\geq c_k(\alpha)/2$. If instead
$\Delta p^2 < c_k(\alpha)/2$ the mode is unstable and grows with the rate~Re$(\lambda_k)$.

\section{Numerical simulations}
\label{simulations}

We shall now compare the predictions of the previous Section concerning the stability
of the waterbag initial state for the $\alpha$-HMF model with numerical simulations. 
We integrate the equations of
motion (\ref{eq:qj},\ref{eq:pj}) using an optimized fourth-order symplectic scheme \cite{atela}
with time step $0.1$, which guarantees a good conservation of the energy during time evolution.
We do not need here to perform long-time evolutions, because we have just to look at the initial
growth of the Fourier modes. Hence, we can push as much as we can towards larger values of
$N$: we report here simulations with $N$ up to $2^{18}$. The reason for choosing a power of
$2$ for N is that we use a Fast Fourier Transform algorithm in order to compute the force
in Fourier space: this allows us to perform simulations in $O(N \ln N)$ time instead
of the $O(N^2)$ time needed to compute the double sum over the sites of the lattice.
We recall that for the $\alpha$-HMF model $v(q_j,q_k)=-\cos (q_j -q_k)$ and we choose here
$\alpha=0.8$.

\begin{figure}[!ht]
\centerline{
$\begin{array}{c}
\epsfig{figure=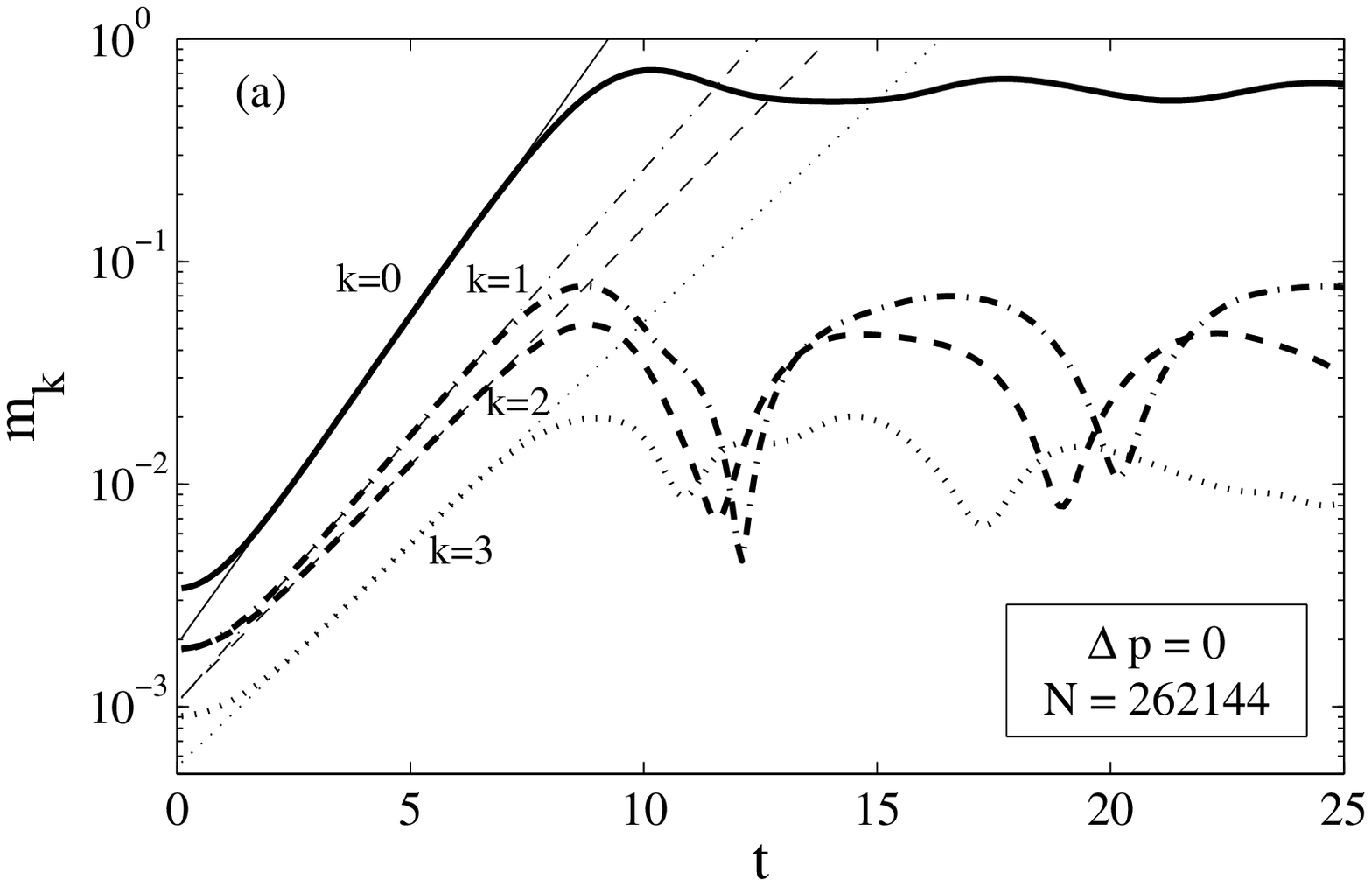,width=9cm}
\\ \epsfig{figure=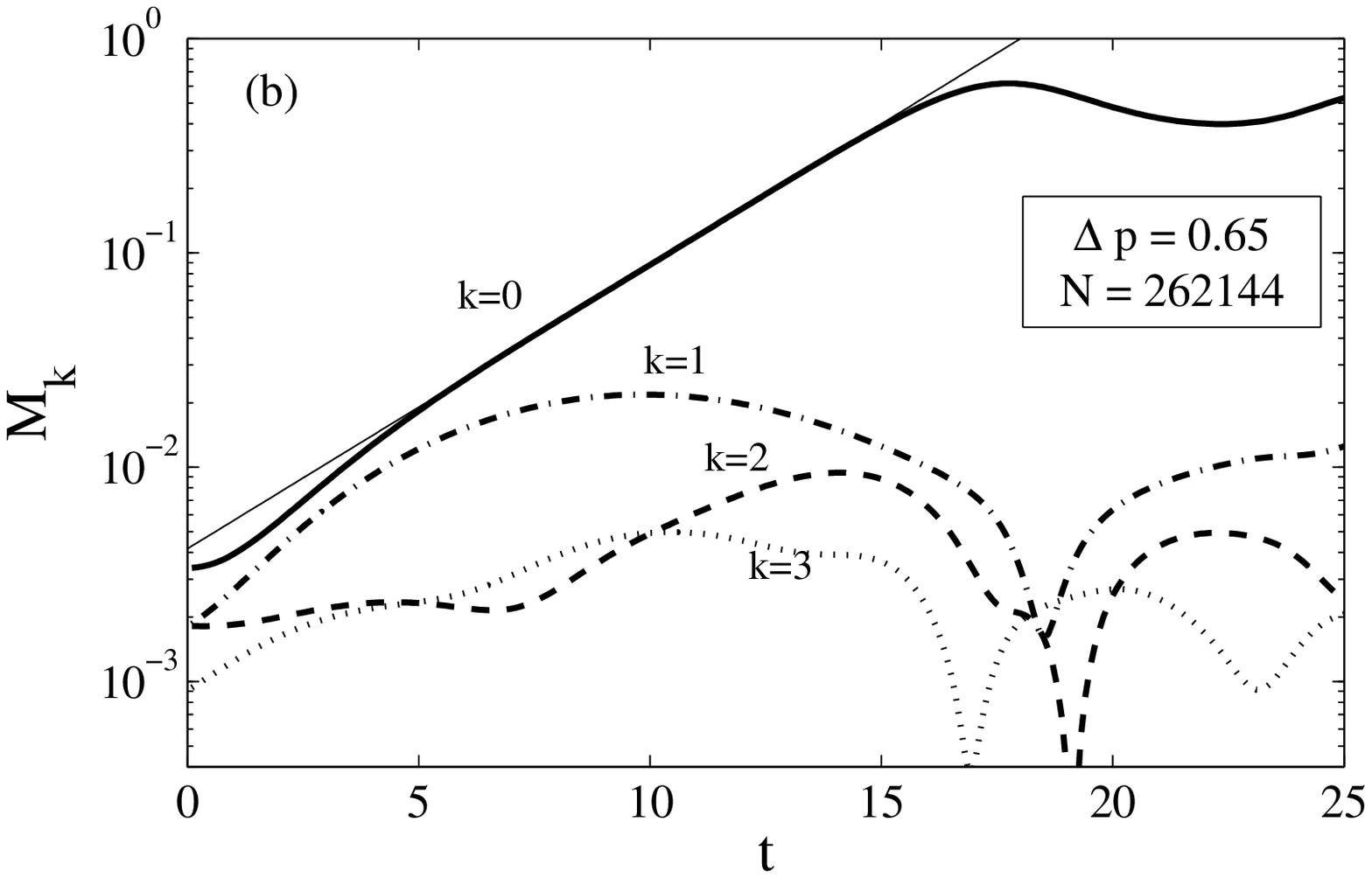,width=9cm}
\\ \epsfig{figure=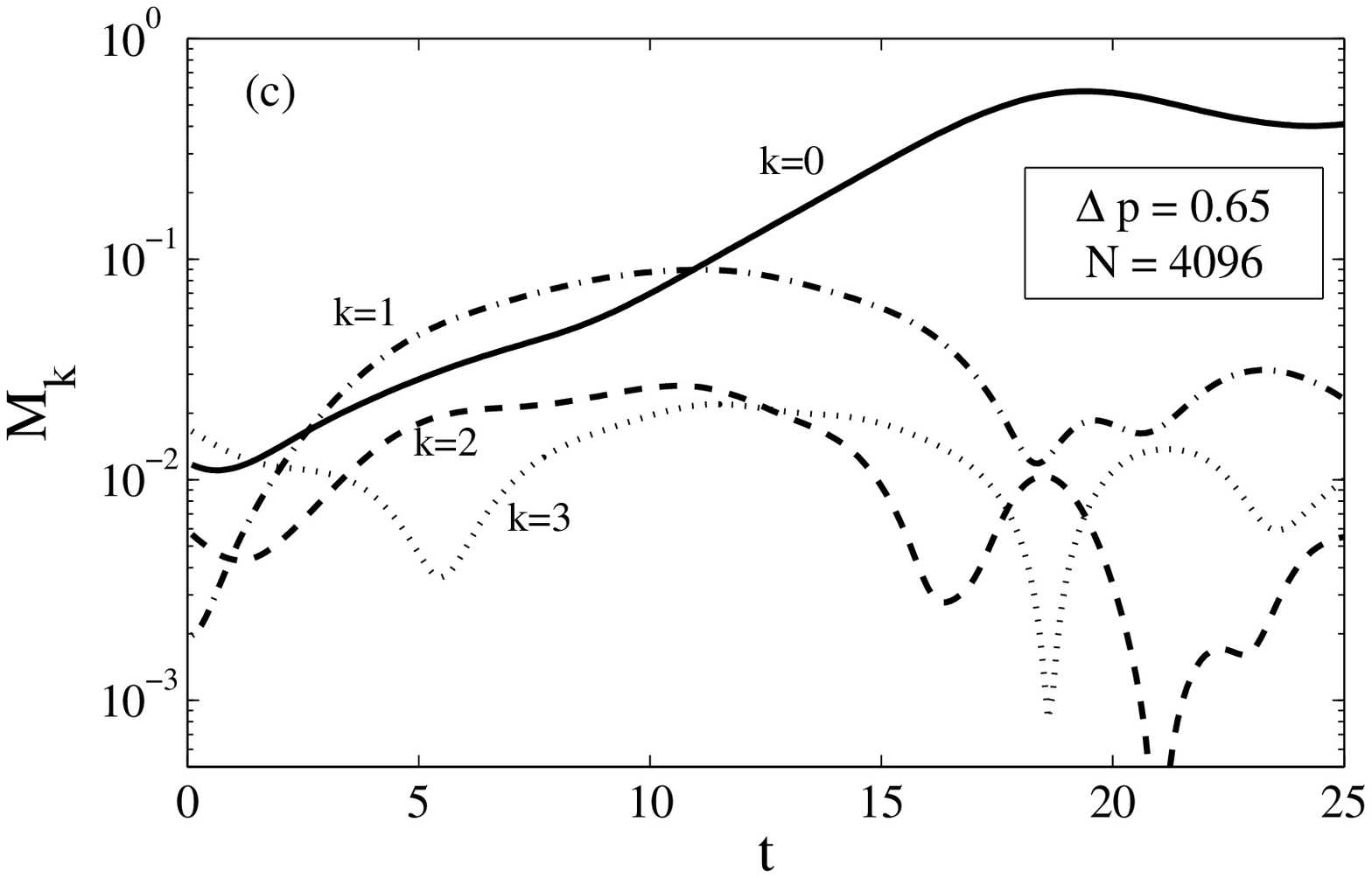,width=9cm}
\end{array}$}
\caption{Time evolution of the Fourier modes of the magnetization $m_k(t)$ (\ref{moments}) for the 
$\alpha$-HMF model. The initial waterbag distribution is uniform in $q \in [0,2\pi]$ and in $p \in [-\Delta p,\Delta p]$
with $\Delta p=0$ in panel (a) and $\Delta p=0.65$ in panels (b) and (c). 
For these values of $\Delta p$ the homogeneous waterbag is unstable. The Fourier modes $m_k$ are plotted 
vs. time for $k=0$ (thick full line), $k=1$ (thick dash-dotted line), $k=2$ (thick dashed line) and $k=3$
(thick dotted line). The thin lines are exponential fits.
The two upper panels have $N=2^{18}$, while the lower panel has $N=2^{12}$ and shows the strong finite-size effects: 
the exponential growth is no more clearly visible.
\label{fig:HMFhomo}}
\end{figure}

The system is initiated by spreading the particles randomly in the rectangle $[0;2\pi]\times[-\Delta p;\Delta p]$
in the phase space $(q,p)$ attached to each site of the lattice. The time evolution of each harmonic of the magnetization 
is followed using the observable
\begin{eqnarray}
m_k&=&\left|\iint e^{-ikx}e^{iq}f(q,p;x,t)\dd q\ \dd p\right|
\\ &=&\frac{1}{N}\left|\sum_j e^{-i k x_j} e^{iq_j}\right|.
\label{moments}
\end{eqnarray}
The simulations allow us to observe an exponential growth of $m_k(t)$
(see Fig.~\ref{fig:HMFhomo}(a,b)), with a rate that depends on the harmonic number $k$, as expected 
from formula~(\ref{eq:disphomo3}). They also confirm that some higher harmonics may be stable while lower 
ones are unstable: for example, for $\Delta p=0.65$, formula~(\ref{eq:disphomo3}) predicts that Fourier 
modes with $k\geq 1$ should be stable. Simulations indeed reveal that for such a value of $\Delta p$, 
mode $k=3$ does not grow exponentially any more (see Fig.~\ref{fig:HMFhomo}(b)).

Let us give some further informations about the time evolution of $m_k(t)$. The harmonics are initially 
at a level of order $1/\sqrt{N}$, as expected from a random distribution. The very short-time dynamics ($t<2$ for $\Delta p=0$) 
does not allow us to observe the exponential growth, presumably because there is still a strong competition between the 
numerous eigenmodes, be they stable or not. The most unstable mode eventually dominates, and it is observed that 
the exponential growth sets in for the time range $2<t<7$ for $\Delta p=0$, which is almost the same for all the harmonics 
(the growth rate is calculated over this time window). Remark that for $\Delta p=0.65$, in  Fig.~\ref{fig:HMFhomo}(b), although the growth 
is very close to an exponential one for mode $k=0$, it is not for $k=1,2,3$, as predicted by the theory.

Finally, the results of a simulation with much less particles $N=2^{12}$ are drawn in Fig.~\ref{fig:HMFhomo}(c), to 
illustrate the finite-size effects. In this case the Fourier modes start for $t=0$ at a much higher level, so that 
the growth occurs on a much shorter time window. It becomes then very hard to identify the exponential character of 
the growth.

\begin{figure}[!ht]
\centerline{
$\begin{array}{c}
\epsfig{figure=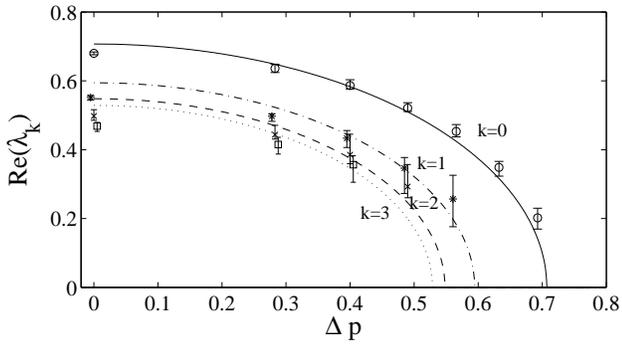,width=9cm}
\end{array}$}
\caption{Growth rates $Re(\lambda_k)$ of the first Fourier modes of the magnetization as a function of $\Delta p$.
The initial conditions are homogeneous waterbags. The lines correspond to the theoretical prediction 
(\ref{eq:disphomo3}) for $k=0$ (full),$k=1$ (dash-dotted), $k=2$ (dashed) and $k=3$ (dotted). The results 
of numerical simulations are marked by the symbols (resp. $\circ$, $\ast$, $\times$ and $\Box$). For each value of
$\Delta p$  ten runs were performed: the symbols represent the average, while the error bars stand for the minimal 
and maximal values. The symbols for $k=1$ and $k=3$ are slightly shifted in $\Delta p$ for the sake of clarity.
\label{fig:graph}}
\end{figure}

In order to test more thoroughly the accuracy of the theoretical predictions, simulations for homogeneous waterbags 
with different values of $\Delta p$ were performed. In Fig.~\ref{fig:graph} we show the growth rates as a
function of $\Delta p$ for different Fourier modes.  The agreement between theory and simulations is excellent 
for the mean-field mode ($k=0$), quite good for the two following modes $k=1,2$ but it gets worse as the harmonic 
number increases, especially for the larger values of $\Delta p$. This phenomenon is possibly due to the fact that 
higher Fourier modes are more sensitive to finite-size effects. Finally, let us comment on the choice $\alpha=0.8$ for our test: 
as can be evaluated from Eq.(\ref{eq:disphomo3}), the domains of instability of $k\geq 1$ modes is very narrow for smaller values of $\alpha$, so that a $\alpha$ close to unity was best to present a stability diagram showing clearly the successive extinction of the Fourier modes. 

\section{Conclusions}
\label{conclusion}

The main result of this paper is the derivation of the Vlasov equation for a lattice with long-range
couplings. As an application, we have studied the stability of the linearized Vlasov equation for 
the $\alpha$-HMF model, deriving a dispersion relation which is expressed in terms of Fourier modes . 

We have computed the stability thresholds for the different Fourier modes and the growth rates when the modes are
unstable. We have shown that the fastest growth rate is the one corresponding to the mean-field
zero mode. This agrees with the widely spread opinion that the $\alpha$-HMF model with $0 \leq \alpha <1$ 
behaves in all respects as the HMF model, which corresponds to the limit $\alpha=0$.

In order to support this conjecture, we have also tried to identify states where higher
harmonics would dominate, performing numerical simulations in which the system is initialized 
in a state which is spatially modulated with a given wave-number in order to promote the time evolution
of higher modes. In all cases, we have observed the growth of other modes, including the zero
mode. This latter eventually takes the lead and dominates over the others.

A more specific remarks concerns the amplitude $c_k(\alpha$ of the Fourier modes of the potential,
given in formula (\ref{cik}). In the limit $\alpha=0$ all the harmonics $k>1$ vanish, giving the 
potential in the mean-field limit, which corresponds to the one of the HMF model.
On the other hand, when going toward the limit $\alpha\to 1$ from below, one gets
$c_k(\alpha)=1$ for all $k$'s. This means that all the Fourier modes, and thus all the 
length scales, become equally relevant in this limit. For $\alpha=1$, the series of Fourier modes 
that represents the potential is no more summable and, as a consequence, the potential (\ref{eq:Vxf}) is
ill-defined: it's the signature that one is leaving the long-range region.

\section{Acknowledgements} 

We thank D. Mukamel for suggesting this problem and for collaboration in a first stage
on this project. This work was carried out in part while S.R. was Weston Visiting 
Professor at the Weizmann Institute of Science. This work is also part of the ANR-10-CEXC-010-01,
{\it Chaire d'Excellence} project.


\begin{thebibliography}{99}

\bibitem{Campa} A. Campa, T. Dauxois and S. Ruffo, Phys. Rep., {\bf 480}, 57 (2009).

\bibitem{Leshouches1}  Dauxois T, Ruffo S, Arimondo E, Wilkens M (Eds.), 
{\it Dynamics and Thermodynamics of Systems with Long-Range Interactions}, 
{\it Lecture Notes in Physics} {\bf 602}, Springer (2002).

\bibitem{Assisi} A. Campa, A. Giansanti, G. Morigi  and F. Sylos Labini (Eds.),
{\it Dynamics and Thermodynamics of systems with long range interactions: theory and experiments}, 
{\it AIP Conference proceedings} {\bf 970} (2008).

\bibitem{Leshouches2} T. Dauxois, S. Ruffo and L. Cugliandolo (Eds.), 
{\it Long-Range Interacting Systems}, {\it Lecture Notes of the Les Houches Summer School: 
Volume 90, August 2008}, Oxford University Press (2009).

\bibitem{Gupta} F. Bouchet, S. Gupta and D. Mukamel, Physica A, {\bf 389}, 4389 (2010).

\bibitem{Barre} J. Barr\'e, D. Mukamel and S. Ruffo, Phys. Rev. Lett., {\bf 87}, 030601 (2001) .

\bibitem{Balescu} R. Balescu, {\it Statistical Dynamics: Matter out of Equilibrium}, 
Imperial College Press, London (1997). 

\bibitem{Nicholson} D. R. Nicholson, {\em Introduction to Plasma Theory}, John Wiley (1983).

\bibitem{yamaguchi} Y. Y. Yamaguchi, J. Barr\'e, F. Bouchet, T. Dauxois and S. Ruffo, 
Physica A, {\bf 337}, 36 (2004).

\bibitem{Henon} M. H\'enon, {Annales d'Astrophysique}, {\bf 27}, 83 (1964).

\bibitem{Lyndenbell} D. Lynden-Bell, Monthly Notices of the Royal Astronomical Society, {\bf 136}, 101 (1967).

\bibitem{dyson1} F. J. Dyson, Commun. Math. Phys. {\bf 12}, 91 (1969).

\bibitem{dyson2} F. J. Dyson, Commun. Math. Phys. {\bf 12}, 212 (1969).

\bibitem{thouless} D. J. Thouless, Phys. Rev. {\bf 187}, 732733 (1969).

\bibitem{anderson} P. W. Anderson and G. Yuval, Phys. Rev. Lett. {\bf 23}, 8992 (1969).

\bibitem{hamann} D. R. Hamann, Phys. Rev. Lett. {\bf 23}, 9598 (1969).

\bibitem{kac} M. Kac, G. E. Uhlenbeck and P. C. Hemmer, J. of Math. Phys., {\bf 4}, 216 (1963).

\bibitem{Tsallis} C. Anteneodo and C. Tsallis, Phys. Rev. Lett., {\bf 80}, 5313 (1998).

\bibitem{tamarit} F. Tamarit and C. Anteneodo, Phys. Rev. Lett., {\bf 84}, 208 (2000).

\bibitem{campa00} A. Campa, A. Giansanti and D. Moroni, Phys. Rev. E {\bf 62}, 303 (2000).

\bibitem{campa03} A. Campa, A. Giansanti, and D. Moroni, J. Phys. A, {\bf 36}, 6897 (2003).

\bibitem{inagaki} S. Inagaki S and T. Konishi, Publ. Astron. Soc. Jpn., {\bf 45}, 733 (1993). 

\bibitem{pichon} Pichon C, 1994 PhD Thesis Cambridge.

\bibitem{hmf95} M. Antoni M and S. Ruffo, Phys. Rev. E, {\bf 52}, 2361 (1995).

\bibitem{vandenberg} T. L. Van Den Berg, D. Fanelli and X. Leoncini, Europhys. Lett., {\bf 89}, 50010 (2010).

\bibitem{turchi} A. Turchi, D. Fanelli and X. Leoncini, arXiv:1007.2065 (2010).

\bibitem{Mori} T. Mori, Phys. Rev. E, {\bf 82}, 060103(R) (2010).

\bibitem{BinneyTremaine} J. Binney J and S. Tremaine, {\it Galactic Dynamics}, Princeton Series in 
Astrophysics (1987).

\bibitem{bertin} G. Bertin, F. Pegoraro, F. Rubini F and E. Vesperini, Astrophys. J., {\bf 434}, 94 (1994).

\bibitem{kaljnas} A. J. Kalnajs, Astrophys. J., {\bf 166}, 275 (1971)

\bibitem{camporeale} E. Camporeale, G. L. Delzanno, G. Lapenta and W. Daughton W, 
Phys. Plasmas, {\bf 13}, 092110 (2006).

\bibitem{lin2001} Z. Lin, Math. Research Lett., {\bf 8}, 1 (2001).

\bibitem{chavanis07} P.-H. Chavanis, Physica A, {\bf 377}, 469 (2007).

\bibitem{us10} R.~Bachelard, F.~Staniscia, T. Dauxois, G. De Ninno, S.~Ruffo,
``Stability of inhomogeneous states in mean-field models with a local potential'', arXiv:1010.4647 
(2010)

\bibitem{braun} W. Braun and K. Hepp, Commun. Math. Phys., {\bf 56}, 101 (1977).

\bibitem{neunzert} H. Neunzert, Fluid. Dyn. Trans., {\bf 9}, 229 (1978).

\bibitem{barre10} J. Barr\'e, A. Olivetti and Y. Y. Yamaguchi, {\it J. Stat. Mech.}, P08002 (2010).

\bibitem{atela}  R. I. Mclachlan and P. Atela, Nonlinearity, {\bf 5}, 541 (1992).

\end{thebibliography}
\end{document}